\documentclass[%
 aip,
 jmp,
 amsmath,amssymb,
 reprint,%
]{revtex4-1}
\usepackage{natbib}
\usepackage{amsfonts}
\usepackage[mathscr]{eucal}
\usepackage{graphicx}
\usepackage{dcolumn}
\usepackage{color}
\usepackage{yfonts}
\usepackage{bm}
\usepackage[inline]{}
\usepackage{ulem} 
\usepackage{xcolor}
\usepackage{booktabs}
\usepackage[utf8]{inputenc}
\usepackage[T1]{fontenc}
\usepackage{mathptmx}
\usepackage{etoolbox}
\newcommand{\kB}{k_\mathrm{B}} 
\newcommand{\Tcs}{T_\mathrm{c}^{\mathrm{s}}} 
\newcommand{\FMF}{F} 
\newcommand{\FC}{F_\mathrm{C}} 
\newcommand{\FCmix}{F_\mathrm{col}} 
\newcommand{\FAB}{F_\mathrm{AB}} 
\newcommand{\NC}{N_\mathrm{C}} 
\newcommand{\NB}{N_\mathrm{B}} 
\newcommand{\NA}{N_\mathrm{A}} 
\newcommand{\VC}{V_\mathrm{C}} 
\newcommand{\EXHS}{\psi_\text{HS}^{\text{ex}}} %
\newcommand{\NCi}{N_{\mathrm{C}_{1}}} 
\newcommand{\NCii}{N_{\mathrm{C}_{2}}} 
\newcommand{\Ndiff}{N_\mathrm{diff}} 
\newcommand{\mc}{ \Delta\mu_{\mathrm{C}}}  
\newcommand{\ms}{ \Delta\mu_{\mathrm{s}}} 
\makeatletter
\def\@email#1#2{%
 \endgroup
 \patchcmd{\titleblock@produce}
  {\frontmatter@RRAPformat}
  {\frontmatter@RRAPformat{\produce@RRAP{*#1\href{mailto:#2}{#2}}}\frontmatter@RRAPformat}
  {}{}
}%
\makeatother
\begin{document}


\title[Paper title]{Binary colloidal mixtures in near--critical binary solvents}
\author{Nima Farahmand Bafi}
 \affiliation{Institute of Physical Chemistry, Polish Academy of Sciences, Kasprzaka 44/52, PL-01-224 Warsaw, Poland}
\author{Robert Evans}
\affiliation{H. H. Wills Physics Laboratory, University of Bristol, Royal Fort, Bristol BS8 1TL, United Kingdom}
\author{Anna Macio\l ek}
\email{amaciolek@ichf.edu.pl}
\affiliation{Institute of Physical Chemistry, Polish Academy of Sciences, Kasprzaka 44/52, PL-01-224 Warsaw, Poland}
\affiliation{Max-Planck-Institut f\"ur Intelligente Systeme, Heisenbergstr. 3, D-70569 Stuttgart, Germany}

\date{\today}

\title{Binary colloidal mixtures in near--critical binary solvents}
\begin{abstract}
The phase behavior of a single type of colloid C suspended in near-critical solvents is known to be very rich. Motivated in part by recent experiments\cite{Kodger-et:2025}  we consider a  mixture of two colloidal types C$_1$ and C$_2$ in a binary solvent close to its demixing critical point. We extend a mean-field description of a lattice model, previously used to investigate systems with a single type of colloid in two dimensions, to the binary colloid case in three dimensions. The model treats the system as a full four-component mixture. For simplicity we choose  C$_1$ and C$_2$ to be hard spheres with the same radius but with different affinities for one species, B, of the AB binary solvent. We show that intricate interplay between couplings of  C$_1$ and solvent, C$_2$ and solvent as well as solvent - solvent interactions and hard sphere packing drive significant changes in the topology of the colloidal phase diagram when the relative volume fractions of the two different colloid types change. The behavior of the two lines of triple points is particularly interesting. Our results can provide some insight into the control of the self-assembly process for colloidal 'alloys' mediated by a near-critical solvent and therefore controlled by temperature in a reversible manner.
\end{abstract}

\maketitle

\section{Introduction}
\label{1-introduction}
Mixtures of colloidal particles, in which two or more distinct types differ in size, shape, or surface chemistry, are particularly valuable from the point of view of potential applications, e.g., in the fabrication of optical~\cite{doi:10.1021/acs.chemrev.6b00196}, catalytic~\cite{doi:10.1021/ja3097527} and plasmonic materials~\cite{doi:10.1073/pnas.1422649112}, but also from a more fundamental viewpoint: these can mimic the behavior of atomic and molecular mixtures and even metallic alloys. Early studies go back to ~\cite{Pusey1989,Bartlett1992}. By adjusting size ratio, concentration, or modifying effective particle interactions via specific surface chemistry, colloidal mixtures can exhibit phase transitions analogous to solid-to-solid alloy transitions, liquid-liquid separation, or eutectic behavior~\cite{Leunissen2005,Solomon2010}. Because of their tunability, colloidal mixtures have been studied extensively in both equilibrium and non-equilibrium contexts, revealing phase diagrams and kinetic pathways that often parallel those of atomic systems but now at accessible length and time scales.

A broad range of experimental techniques has been used to probe these systems, from static and dynamic light scattering \cite{Pusey1989,Bartlett1992} to advanced confocal microscopy \cite{Leunissen2005,Solomon2010,Lu2008}. Experiments have demonstrated phase separation in size-asymmetric mixtures \cite{Leunissen2005}, crystallization in binary hard-sphere-like systems \cite{Bartlett1992}, and complex assembly pathways in mixtures with tunable attractions \cite{Solomon2010}. These pioneering investigations highlight how colloidal mixtures can serve as analogues for probing fundamental aspects of multicomponent phase behavior.

The effective interaction between colloids depends on the surrounding solvent. So called solvent-mediated (SM) interactions arise when the presence of a colloid perturbs the local structure or composition of the solvent, thereby affecting nearby colloidal particles. This gives rise to a potential of mean force between two colloids that depends upon the colloid types and their adsorption properties. In binary solvents, preferential adsorption of one component at the colloid surface can lead to a range of adsorption and wetting phenomena \cite{Floter1995}. 

In the present paper we focus on the situation where the binary solvent is close to the critical point of demixing. Here the range of SM interactions can become very large due to the diverging correlation length $\xi$ of the solvent order parameter. Fisher and de Gennes \cite{Fisher1978} first predicted that confining critical fluctuations between surfaces should lead to a long-ranged, universal force. Later it was pointed out that such forces are analogous to the well-known quantum Casimir effect and these long ranged forces are now termed critical Casimir forces. Of course, these are purely classical in origin. Their nature depends sensitively on the boundary conditions imposed by the colloid surfaces: identical surface preferences lead to attraction, while surfaces with opposite preferences lead to repulsion.

There is ample experimental evidence that critical Casimir forces can be used to tune directly the interaction  between a colloid and a substrate~\cite{schmidt2023tunable,D0NR09076J,wang2024nanoalignment,10.1063/5.0235449} or between a pair of
colloids~\cite{Hertlein-et:2008,Bonn-et:2010a,Gambassi-et:2009,Marcel-et,maciolek2018collective}.
Reversible self-assembly of colloids directed by a
substrate~\cite{Soyka-et:2008,Troendle-et:2011} or
occurring without a template~\cite{Beysens-et:1985,MARINO2016154,Iwashita-et:2014,Marcel-et} were investigated. Implementing Casimir effect-driven assembly has been also successful for probing colloidal gas, liquid or solid phases~\cite{Guo-et:2008,Bonn-et:2009,Nguyen-et:2013}. In this context, the suspensions studied so far are mainly ternary mixtures, i.e. a single type of colloid in a binary solvent close to its critical point of de-mixing.  Much less is known about binary colloidal mixtures in equivalent critical solvents. Reference~\cite{Zvyagolskaya-et:2011} investigated
phase separation for a binary colloidal suspension where the colloidal types have opposite adsorption preferences for the components of a binary solvent.  More recently, experiments investigated colloidal phase behavior for a binary colloidal mixture in which the two colloidal types prefer, to a different degree,  the same component of a binary solvent~\cite{Kodger-et:2025}. This study prompted us to ask a basic question: what new physics, meaning features of phase diagrams, might emerge for a binary colloidal mixture additional to those present for a single colloidal type?

Existing theoretical treatments, for the case of a single colloidal type, fall into two broad classes. First are effective one-component approaches, in which the solvent is integrated out and its influence captured by a state-dependent effective pair potential acting between two colloids. Such effective potentials can be constructed from simulations that determine critical Casimir scaling functions \cite{Vasilyev2009,Mohry-et:2012a,Mohry-et:2012b,Gnan-et:2012c,Mohry-et:2014} or fitted to experimental force measurements \cite{Dang-et:2013,Nguyen-et:2013,Marcel-et}. Such approaches are certainly effective in describing dilute suspensions. However, these might fail to capture many-body SM interactions that become important in dense systems, see e.g.the review \cite{maciolek2018collective}.
The second class comprises explicit multicomponent mixture models, in which both solvent species and colloids are retained as distinct components and treated on essentially equal footing. These can account naturally for many-body effects and critical point shifts; early attempts can be found in \cite{Sluckin:1990,Loewen:1995,GIL1998245}. Finding a model that permits tractable simulations of large colloids immersed in in a binary molecular solvent is very challenging. The lattice-based ternary ABC model investigated by Edison et al. \cite{Edison2015PRL,10.1063/1.4961437,edison2015}, in which colloids C are hard disks embedded in a two dimensional (2D) binary lattice gas solvent, consisting of species A and B with species B preferentially adsorbed on the colloids via a surface energy term, has much appeal. Monte Carlo simulations in 2D~\cite{Edison2015PRL,10.1063/1.4961437} revealed a rich variety of phase transitions, i.e. gas-liquid (GL), gas–solid (GS) and solid-solid (SS), as well as significant shifts in the ternary critical point relative to that of the colloid - free solvent. A later Monte Carlo study in 3D ~\cite{Tasios-et:2017}, where the hard disks are replaced by hard spheres and the solvent lives on a simple cubic lattice, showed that the topology of the phase diagrams was close to that found in 2D. The 3D simulation was a computational tour de force. Importantly the simulation results reinforce the need to develop approximate theories for determining phase phase behavior. A mean-field theory \cite{edison2015} for the 2D model captured qualitatively most features of the simulation phase diagrams and it is this treatment and the physical insight provided that we build upon here.

Specifically, we extend the mean-field framework of Ref.~\cite{edison2015} by adding a second colloidal type. We investigate a four component system with two colloidal types and work in 3D. The underlying Hamiltonian is that in ~\cite{Tasios-et:2017} straightforwardly extended to two colloidal types. In order to keep the parameter space manageable we set the radius of each colloid type to be the same. Of course, this involves replacing the 2D hard-disk equation of state by the corresponding 3D hard-sphere form \cite{carnahan1969,Speedy1998} and adjusting coordination numbers and excluded volumes to 3D geometry. Our formulation retains explicit solvent degrees of freedom, enabling the study of many-body effects and the coupling between solvent criticality and binary colloidal phase behavior. As far as possible, we retain the notation of the earlier papers. By combining the minimal lattice model with a three-dimensional mean-field description, we aim to provide a simple but  physically well - grounded framework for interpreting experimental results on the phase behavior of binary mixtures of colloidal particles suspended in near-critical solvents.

Our paper is organized as follows: In Sec. IIA we introduce the model and the approximate free energy that we utilize. This subsection relies upon~\cite{edison2015}. Sec. IIB presents expressions for the chemical potentials and pressure while Sec. IIC describes how phase coexistence and critical points are determined. Results are presented in Sec. III. We display phase diagrams for the limiting case of a ternary mixture containing only type 1 colloids in Sec. IIIA - see figures 1-4. Phase diagrams for the binary colloidal system, a four component mixture, are given in Sec. IIIB - see figures 5-11. We conclude in Sec. IV with a summary of our results and their relevance for recent experiments~\cite{Kodger-et:2025}. The appendix describes the approximation we have employed for the free energy of a pure hard sphere colloidal system.
\section{Theory}
\label{sec:theory}
\subsection{Description of Model and the Approximate Free energy.}
\label{subsec:model}
\begin{table*}[htbp]
\caption{Definitions of symbols used in the model.}
\label{definitions}
\centering
\begin{ruledtabular}
\begin{tabular}{lc} 
\toprule
Symbol & Description \\
\midrule
\hline
$M$ & Number of lattice sites (volume in units of volume of lattice site) \\
$\NCi$ & Number of colloidal type 1 \\
$\NCii$ & Number of colloidal type 2 \\
$\NC := \NCi + \NCii$ & Total number of colloids \\
$\Ndiff := \NCii - \NCi$ & Difference of the number of colloids \\
$\NB$ & Number of molecules of solvent species B \\
$\eta_1$ & Fraction of $M$ occupied by colloids type 1 \\
$\eta_2$ & Fraction of $M$ occupied by colloids type 2 \\
$\eta := \eta_1 + \eta_2$ & Fraction of $M$ occupied by all colloids \\
$\delta := \eta_2 - \eta_1$ & Difference in colloid fractions \\
$M(1-\eta)$ & Free space for the solvent without colloids \\
$x$ & Fraction of total sites occupied by B \\
 $1-\eta - x$ & Fraction of total sites occupied by A \\
 $\epsilon>0$ & Strength of AB pair interaction \\
  $\alpha_1$ & Adsorption strength of species B around colloid 1\\
  $\alpha_2$ & Adsorption strength of species B around colloid 2\\
$x_\mathrm{r}$ & Composition in solvent reservoir\\
F, G, L, S & Fluid, gas, liquid, and solid phases, respectively\\
\bottomrule
\end{tabular}
\end{ruledtabular}
\end{table*}
In this section we extend the model introduced in Ref.~\cite{edison2015},  to study the phase diagram of a mixture of two colloidal types C$_1$ and C$_2$ in a binary liquid mixture (solvent) of species A and B. Our model now lives in 3D and the colloids are hard spheres (HS) .Both types have the same radius $R$ but adsorb species B with different strength. The mean--field Helmholtz free energy of this system can be written in terms of the variables defined in table~\ref{definitions}:
\begin{equation}
 \label{F-proposed}
 \begin{split}
\FMF(\NCi,\NCii,M,T,\NB)=&\FCmix(\NCi,\NCii,M,T)\\
&+(1-\eta)\FAB(\NB,M,T)\\
&+\NCi U_\text{BC$_1$}+\NCii U_\text{BC$_2$}.
\end{split}
\end{equation}
The notation builds upon that in Ref.~\cite{edison2015}:
$\FAB$ is the free energy of the solvent,  $U_\text{BC$_i$}$ with $i=1,2$ describes the interaction of a colloid of type $i$ with B particles, and $\FCmix$ is the free energy of the colloidal binary hard sphere mixture without solvent present and is given by
\begin{equation}
 \label{FCmix}
\begin{split}
  \FCmix(\NCi,\NCii,M,T)=&  \FC(\NCi+\NCii,M,T)\\
  &+F_{\mathrm{C}_1\mathrm{C}_2}(\NCi,\NCii,T).  
\end{split}
\end{equation}
$M$ is the total number of lattice sites and we introduce $\eta$ as the fraction of the sites occupied by the colloids.
We approximate $\FC$ using results for a pure hard sphere system that takes into account the values of the melting and freezing packing fractions determined from Monte Carlo simulations~\cite{robles2014,nayhouse2011}. The explicit form used for $\FC$ is summarized in Appendix~\ref{sec: pure-colloids-calculations} and is based on standard references~\cite{carnahan1969,hall1972}.  In the fluid (F) phase it is given by:
\begin{equation}
\label{F-fluid-In-Text}
\begin{split}
\frac{\FC^\text{F}}{M}=\kB T\big(\frac{\eta}{\VC}\ln\Lambda^3\frac{\eta}{\VC}-\frac{\eta}{\VC}\big)+\kB T\frac{\eta^2(4-3\eta)}{\VC(1-\eta)^2},
\end{split}
\end{equation}
where in the ideal gas piece we take $\Lambda^3=\VC$, with $\VC=4/3\pi R^3$ denoting the volume of the colloid so that $\eta=(\NCi+\NCii)\VC/M$. The free energy in the solid (S) phase is given by:
\begin{equation}
\label{F-solid-In-Text}
\begin{split}
\FC^\text{S}/M=&\frac{\kB T}{\VC}\eta\Big[6451.95+2248.99\,\ln\eta\\
&-3\,\ln(0.740489-\eta)-20548.6\,\eta
\\
&+39141.7\,\eta^2-52967.3\,\eta^3\\
&45297.6\,\eta^4 - 22003.8\,\eta^5 + 4631.26\,\eta^6\Big].
\end{split}
\end{equation}
$F_{\mathrm{C}_1\mathrm{C}_2}$ is the (ideal) entropic contribution due to the mixing of the two colloidal types and is given by:
\begin{equation}
    F_{\mathrm{C}_1\mathrm{C}_2}=\kB T\left(\NCi \ln\frac{\NCi}{\NC} +\NCii \ln\frac{\NCii}{\NC}\right).
\end{equation}
Note, that i) $F_{\mathrm{C}_1\mathrm{C}_2}$ is not a function of $x$, the fraction of total sites occupied by species B and ii) that the term $\FC$ is a function of only the sum of colloid numbers. The contribution  $(1-\eta)\FAB$ in eq.~(\ref{F-proposed}) is the mean-field free energy of the binary solvent in the free space between the colloids. It is given by
\begin{equation}
\label{F-solvent}
\begin{split}
\frac{\FAB}{M}=
&Z_\text{ss}\frac{\epsilon}{2}\frac{x(1-x-\eta)}{1-\eta}\\
&+kT\Big[x\ln(\frac{x}{1-\eta})+(1-x-\eta)\ln\left(\frac{1-x-\eta}{1-\eta}\right) \Big],
 \end{split}
\end{equation}
where $Z_\text{ss}$ is the coordination number of the lattice; for the FCC lattice  this is 12.  $\epsilon$ is the energy of nearest neighbor AB interactions. The second term in eq.~(\ref{F-solvent}) is entropic and mimics that in Ref.~\cite{edison2015}.

The solvent-colloid $C_i$  interaction terms are given by:
\begin{equation}
 \label{UBC-two-component}
 \begin{split}
&\NCi U_\text{BC$_1$}= -\NCi Z_\text{cs} \alpha_1 \epsilon \frac{x}{1-\eta}\\
&\NCii U_\text{BC$_2$}= -\NCii Z_\text{cs} \alpha_2 \epsilon \frac{x}{1-\eta}
\end{split}
\end{equation}
where $-\alpha_i\epsilon$ , with $\alpha_i>0$ and $i=1,2$ ,  denotes the energy reduction arising from adsorption of a species B solvent molecule by a colloid of type $i$ and $Z_\text{cs}=4\pi R^2$ (recall  $R$ is the radius of both colloidal types) is the total number of B molecules that fit to the surface of the colloids.
Finally, the free energy per site can be written as
\begin{equation}
 \label{F-Final}
 \begin{split}
\frac{\FMF}{M}=&-\frac{\eta}{2\VC}Z_\text{cs} (\alpha_1+\alpha_2) \epsilon \frac{x}{1-\eta}
\\&-\frac{\delta}{2\VC}Z_\text{cs} (\alpha_2-\alpha_1) \epsilon \frac{x}{1-\eta}\\
&+Z_\text{ss}\frac{\epsilon}{2}\frac{x(1-x-\eta)}{1-\eta}\\
&+\kB T\Big[x\ln(\frac{x}{1-\eta})+(1-x-\eta)\ln(\frac{1-x-\eta}{1-\eta}) \Big]\\
&+\FC^{K}/M\\
&+\frac{\kB T}{2\VC}\Big[(\eta-\delta)\ln(\frac{\eta-\delta}{\eta})+(\eta+\delta)\ln(\frac{\eta+\delta}{\eta}) \Big],
 \end{split}
\end{equation}
where we  recall
$\NC/M=\eta/\VC$, $\NCi/M=\eta_1/\VC$, $\NCii/M=\eta_2/\VC$,  $\eta=\eta_1+\eta_2$ and $\delta=\eta_2-\eta_1$. 
Here, $\FC^K$  with $K$= F or S, denotes the colloid contribution to the free energy, which depends on whether the system is in the fluid phase ($\FC^K=\FC^\text{F}$) or it is in the solid phase ($\FC^K=\FC^\text{S}$).
Depending on the representation of the results we  can exchange the variables from $(\eta,\delta,x)$ back to $(\eta_1,\eta_2,x)$ in eq.~(\ref{F-Final}). Note that eq.~(\ref{F-Final}) reduces to eq.~(4) in Ref.~\cite{edison2015} in the pure colloid limit where $\NCii=0$, $\delta=0$ and $\alpha_1=\alpha_2$ However, we now consider a $3$ dimensional system.
\subsection{Chemical potentials and pressure}
\label{subsec:chem_pot}

The total free energy in our mean-field approximation is a function of the variables $(\NCi,\NCii,M,T,\NA,\NB)$, where $M$ plays the role of volume. Therefore
\begin{equation}
\label{differentials1}
\begin{split}
        \mathrm{d}\FMF =& -S \mathrm{d}T - p \mathrm{d}M+ \mu_ \mathrm{C_1} \mathrm{d}\NCi+\mu_\mathrm{C_2} \mathrm{d}\NCii
        \\
        &
        +\mu_\mathrm{A} \mathrm{d}\NA
        + \mu_\mathrm{B} \mathrm{d}\NB,
\end{split}
\end{equation}
where $p$ corresponds to a pressure.
Noting that $M=\NA+\NB +\VC\NCi+\VC\NCii=\NA+\NB +V_C\NC$ and using $\NC=\NCi+\NCii$ and $\Ndiff=\NCii-\NCi$, we can  rewrite eq.~(\ref{differentials1}) as
\begin{equation}
\label{differentials4}
\begin{split} 
\mathrm{d}\FMF =& -S \mathrm{d}T - p \mathrm{d}M
    -\mu_\mathrm{A} \mathrm{d}M
    + (\mu_\mathrm{B}-\mu_\mathrm{A}) \mathrm{d}\NB  
    \\
    &+(\mu_ \mathrm{C_1}/2+\mu_ \mathrm{C_2}/2-\mu_\mathrm{A} \VC)\mathrm{d}\NC
    \\
    &
    +(\mu_ \mathrm{C_2}/2-\mu_ \mathrm{C_1}/2)\mathrm{d}\Ndiff.
\end{split}
\end{equation}
Due to the incompressible lattice nature of the solvent, $\NB, N_{C_1}$ and $ N_{C_2}$ suffice to describe uniquely the number of particles of each species present in the system. Thus the four-component system is equivalent to a ternary  system. Based on the above, we define the pertinent chemical potentials and pressure  as 
\begin{equation}
 \label{clarify2b}
 \begin{split}
 &\Delta\mu_\text{s}:=\mu_\text{B}-\mu_\text{A},\\
 &\Delta\mu_\text{C}:=\mu_{\text{C}_1}/2+\mu_{\text{C}_2}/2-\VC\mu_\text{A},\\
&\Delta\mu_\text{$\delta$}:=\mu_{\text{C}_2}/2-\mu_{\text{C}_1}/2,\\
&P:=p-\mu_\text{A},
 \end{split}
\end{equation}
which are given by the following partial derivatives
\begin{equation}
 \label{chemical-potentials}
 \begin{split}
 &\Delta\mu_\text{s}=\frac{\partial(\FMF/M)}{\partial x}|_{T,\eta,\delta,M},\\
 &\Delta\mu_\text{C}=\VC\frac{\partial(\FMF/M)}{\partial\eta}|_{T,\delta,x,M},\\
&\Delta\mu_\text{$\delta$}=\VC\frac{\partial(\FMF/M)}{\partial\delta}|_{T,\eta,x,M},
 \end{split}
\end{equation}
and 
\begin{equation}
 \label{pressure}
 P=-\frac{\partial\FMF}{\partial M}|_{T,\NC,\Ndiff,\NB}.
\end{equation}
Note that, when calculating the pressure  \textit{P} we have to consider that $x$, $\eta$ and $\delta$ in eq.~(\ref{F-Final}) all depend on $M$.
\subsection{Phase coexistence and critical points}
\label{subsec:phase_coex}

The coexisting phases are given by the set of states $(\eta^K,\delta^K,x^K)$, where $K$  denotes different phases, that have the same temperature, chemical potentials (given by eqs.~(\ref{chemical-potentials})) and pressure~(eq.~\ref{pressure}).

In order to find the critical points we first define the following matrices~\cite{tester1997thermodynamics}:
\[
W=\begin{bmatrix}
\FMF_{xx} & \FMF_{x\eta_1} & \FMF_{x\eta_2} \\
\FMF_{\eta_1 x} & \FMF_{\eta_1 \eta_1} & \FMF_{\eta_1 \eta_2} \\
\FMF_{\eta_2 x} & \FMF_{\eta_2 \eta_1} & \FMF_{\eta_2 \eta_2}
\end{bmatrix}
\]
and
\[
\tilde{W}=\begin{bmatrix}
\FMF_{xx} & \FMF_{x\eta_1} & \FMF_{x\eta_2} \\
\FMF_{\eta_1 x} & \FMF_{\eta_1 \eta_1} & \FMF_{\eta_1 \eta_2} \\
(\det{W})_x & (\det{W})_{\eta_1} & (\det{W})_{\eta_2}
\end{bmatrix},
\]
where the subscripts of $\FMF$ and $W$ denote their partial derivatives with respect to the variable in their indices.
The condition for the critical surface is given by~\cite{tester1997thermodynamics}
\begin{equation}
\label{eq:crit_cond}
    \det W=\det\tilde{W}=0
\end{equation}
Note that choosing the set of variables $(\eta_1,\eta_2,x)$ or $(\eta,\delta,x)$ does not change the condition for criticality. One could choose one or the other. For $\eta_1=\eta_2=0$ the model reduces to a binary solvent mixture, which has a critical point at $x_{rc}=0.5$ and the mean--field critical temperature (for $Z_\mathrm{ss}=12$)
$\Tcs=3\epsilon$. In the following we present the temperature values in terms of the reduced temperature $\Delta T/\Tcs:=(T-\Tcs)/3\epsilon$.
We focus here on the temperature range $\Delta T>0$, which corresponds to a single mixed phase in the colloid-free solvent. Moreover, we consider the situation where the reservoir containing the binary solvent is rich in species A , i. e., $\Delta \mu_s < 0$. Most experiments investigating the critical Casimir effect take place under such conditions.

\section{Results}
\label{sec:results}

We first present our results for the limiting case of a mixture containing only type 1 colloids, and then consider the case of a mixture containing type 1 and type 2 colloids.
\subsection{Limiting case of a ternary system (one type of colloid)}
\label{subsec:lc}

The model in sec.~\ref{subsec:model} reduces to the case of a mixture containing only type 1 colloids if we set $\NCii$ = 0. In this limit there is no difference between the colloids, $\delta=\eta_2-\eta_1=0$, so the last term in the equation ~(\ref{F-Final}) becomes zero and  the chemical potential $\Delta\mu_\delta$ in  eq.~(\ref{chemical-potentials}) is irrelevant. This case was investigated in $2$D in ref.~\cite{edison2015}. Here we study it in $3$D.

\begin{figure}[htbp]
   \includegraphics[width=\linewidth]{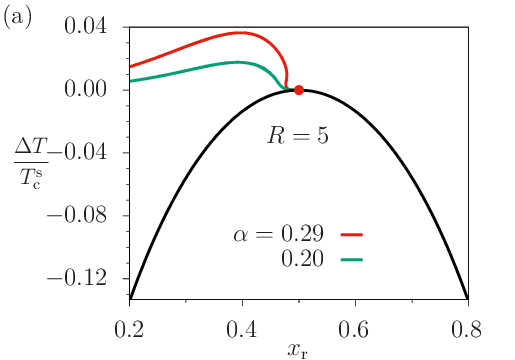}
   \includegraphics[width=\linewidth]{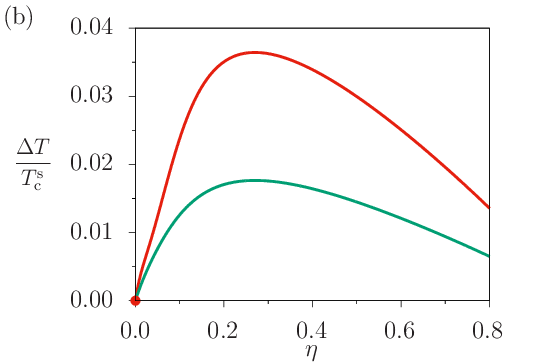}
\caption{(a) Demixing curve of a colloid free solvent (black curve) and loci of critical points of the ternary mixture where only type 1 colloids are added. The critical lines for both choices of $\alpha$ run into the solvent critical point at reservoir composition $x_r$ = 0.5.
(b) Values of $\eta$ along the critical lines in panel (a); the color code is the same. These lines are  solutions to eqs.~(\ref{eq:crit_line}). For larger values of $\eta$ the critical points become metastable with respect to the GS coexistence.
}
\label{fig:Demix-TC-1Type}
\end{figure}

For the ternary mixture, the conditions for the critical points (eqs.~\ref{eq:crit_cond}),  using an obvious notation, reduce to
\begin{equation}
\label{eq:crit_line}
\begin{split}
&\FMF_{\eta\eta}\FMF_{xx} - \FMF_{\eta x}^2= 0,
\\
&\FMF_{xxx}\FMF_{\eta\eta}^2
-\FMF_{\eta\eta\eta}\FMF_{xx}\FMF_{\eta x}
- 3\FMF_{\eta xx}\FMF_{\eta\eta}\FMF_{\eta x} 
\\&
+3\FMF_{\eta\eta x}\FMF_{\eta\eta}\FMF_{xx}= 0.
\end{split}
\end{equation}
The demixing curve for the colloid-free and the loci of critical temperatures upon adding colloid type 1 are plotted versus reservoir concentration in Fig.~\ref{fig:Demix-TC-1Type}(a) for two values of the adsorption strength $\alpha$ and radius $R=5$ (in units of the lattice spacing). The value of the colloid fraction $\eta$ along these critical lines is given in  Fig.~\ref{fig:Demix-TC-1Type}(b). Qualitatively, the behavior of the critical lines is the same as in $2$D. In particular, for larger values of $\alpha$ we observe a non-monotonic approach to the solvent critical point and a shift of the critical line towards higher temperatures.

\begin{figure}[htbp]
    \includegraphics[width=\linewidth]{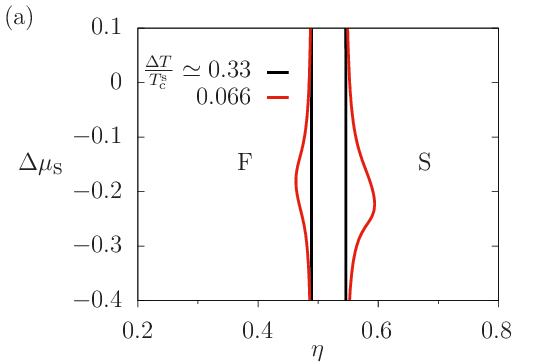}
    \includegraphics[width=\linewidth]{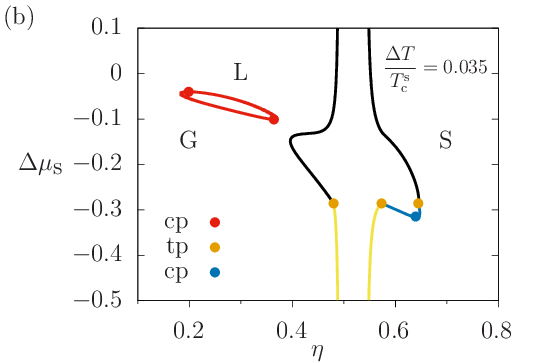}
    \includegraphics[width=\linewidth]{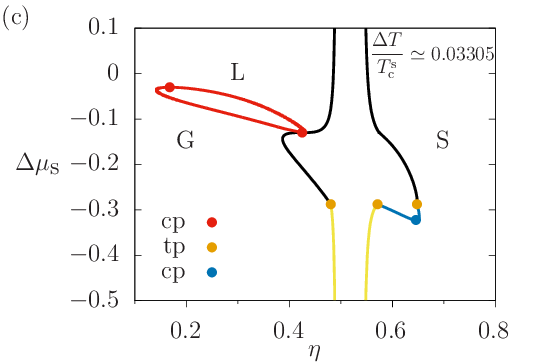}
\caption{ Phase diagrams of a ternary mixture consisting of only type 1 colloids and a binary solvent for different temperatures as listed in the panels. G, L, and S denote the gas, liquid, and solid phases, respectively.
In all cases, $R = 5$ and $\alpha = 0.29$. Panel (a) displays only FS coexistence. New phase boundaries emerge as the temperature is lowered. In panels (b) and (c) a distinct GL coexistence is manifest with upper and lower critical points (cp red dots). The emergence of SS coexistence is signaled by the blue curve with a corresponding critical point (cp blue dots).
The triple GSS point between the G and two S phases is denoted by orange dots.
}
\label{fig:phases-1D-3T}
\end{figure}

In Fig.~\ref{fig:phases-1D-3T} we show how the topology of the phase diagram in the $(\eta, \ms)$ representation changes with decreasing temperature towards $\Tcs$. At high temperatures (panel (a)) the system exhibits only a F and a S phase. Unlike the pure hard sphere system where, of course, the FS phase transition occurs at temperature-independent packing fractions (see Appendix~\ref{sec: pure-colloids-calculations}), the addition of solvent causes a bulging of the coexistence curves as temperature decreases; see red curves. Note that the black lines in panel (a) are not perfectly vertical, although their variation is tiny.
Upon decreasing temperature further, the S
branch of the FS coexistence moves towards higher $\eta$  and a SS transition occurs (see panel (b) of Fig.~\ref{fig:phases-1D-3T}), which corresponds to $\Delta T/T_\mathrm{c}^\mathrm{s}=0.035$. Moreover, the system exhibits a GL transition with an upper and a lower critical point (red points). The orange points denote triple point (GSS) coexistence where a G phase coexists with two S phases with different values of $\eta$. The SS transition is indicated by blue lines and terminates in the (blue) critical point.
\begin{figure*}[t]
\centering
\includegraphics[width=0.82\textwidth]{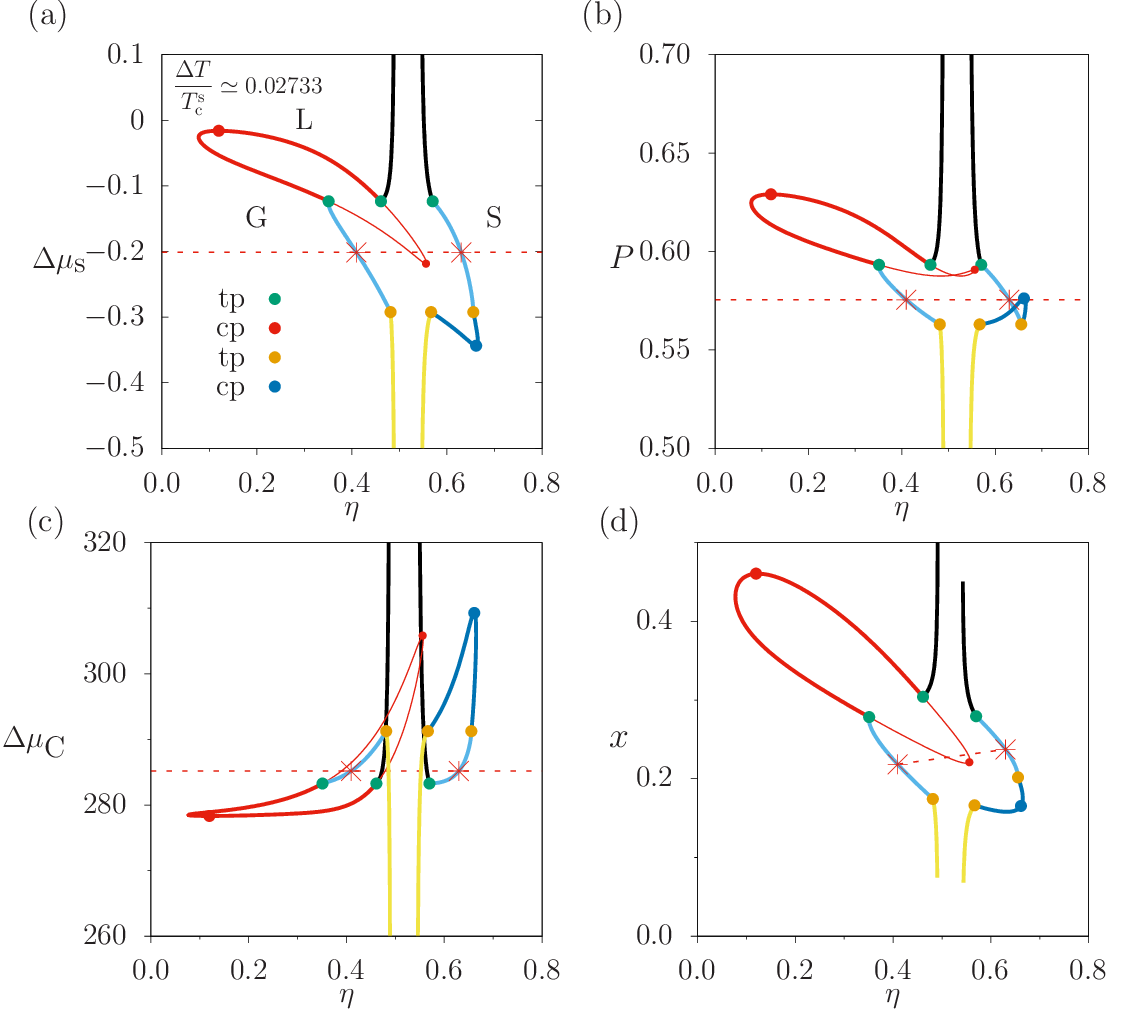}
\caption{ Phase diagram of ternary mixture for $\Delta T/T_\mathrm{c}^\mathrm{s}$=0.02733 for $R=5$ and $\alpha=0.29$ in various representations. The color code is the same in all panels. The red lines denote GL coexistence terminating in a (stable) upper critical (red) point. The thin red lines display the corresponding metastable branches ending in the lower (metastable) critical point. There are two sets of triple points: tp green dots denote the upper GLS coexistence whereas tp orange dots denote the lower GSS coexistence. cp blue denotes the (lower) critical point of SS coexistence. As explained in the text, the red asterisks label GS coexistence. }
\label{fig:low-T-4-fields}
\end{figure*}

Note that along these lines the value of $x$ changes as well which is not shown here. 
If we lower the temperature even further, the lower critical point of the GL transition
shifts towards lower values of $\ms$ and higher values of $\eta$ until it starts to coexist with the S phase at $\Delta T/T_\mathrm{c}^\mathrm{s}\simeq0.03305$, as in panel (c) of Fig.~\ref{fig:phases-1D-3T}.

Moving even closer to $\Tcs$, the GL transition
(red lines) with a lower critical point enters the LS transition region, as shown in Fig.~\ref{fig:low-T-4-fields}(a)-(d) in different representations. This leads to the development of a triple point between the G, L, and S phases.
For solvent chemical potentials below this triple point value, the GL transition
becomes metastable (thin, solid red curves in each panel.)
The red asterisks in the panels of Fig.~\ref{fig:low-T-4-fields} correspond to the coexistence of a G and a S phase.
The values of $\eta$ for these two phases are $\eta=0.4094$ and $\eta=0.6299$ and the chemical potential $\ms=-0.2013$. The values of pressure and colloid chemical potential for these coexisting states can be read from panels (b) and (c). Panel (d) presents results in the $(\eta,x)$ representation where, unlike the other representations, coexistence is not defined by a horizontal line. It should be noted that the full phase diagram should be represented in the space spanned by all independent thermodynamic fields, e.g., $(T, P,\mc,\ms)$ or $(T, P,\eta,x)$. The graphs shown in the different panels are specific cuts of the full representation. As a result, some of the curves in these graphs appear to intersect. However, these intersection points are all distinct states when we consider the values of all thermodynamic fields. The actual intersection points, i.e., triple points, are marked consistently with dots.

\begin{figure}[!ht]
\centering
\includegraphics[width=0.45\textwidth]{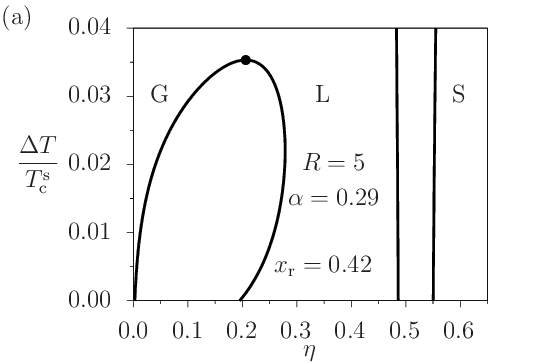}
\includegraphics[width=0.45\textwidth]{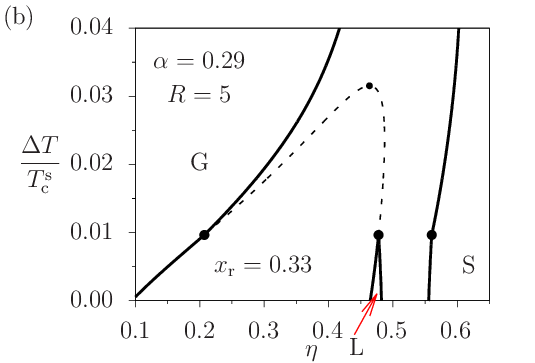}
\includegraphics[width=0.45\textwidth]{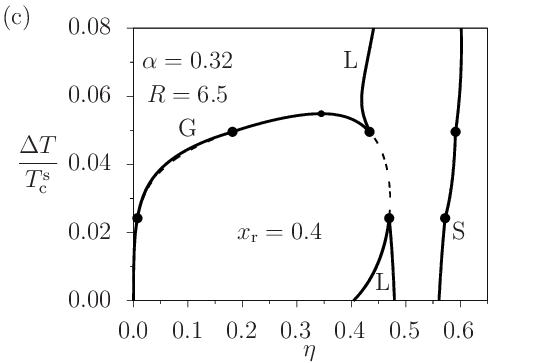}
\caption{ Phase diagram in the $(\eta, \Delta T/\Tcs)$ representation at different values of the solvent reservoir concentration  $x_r$ for radius $R=5$ and adsorption strength $\alpha=0.29$, (a) and (b), and for $R=6.5$ and $\alpha=0.32$ (c). In panel (a) GL coexistence is well separated from LS. In panel (b) GL phase separation becomes metastable -see dashed line and the stable L region
is small. The dots denote the triple GLS line. In panel (c) there is an upper and a lower triple GLS line marked by dots.
}
\label{fig:diff_par}
\end{figure}

It is important to compare phase diagrams obtained from the present MF treatment in 3D with those from the corresponding analysis in 2D (see ref.~\cite{edison2015}). We note many common features. For example, in both dimensions, the ternary mixture exhibits GL and GS coexistence 
with concomitant solvent separation. Moreover, these transitions can occur far from the critical point of the pure solvent, although the line of GL critical points merges smoothly to the critical point of the solvent. In general, we can state that, for a similar (reduced) deviation from  $\Tcs$, in $3$D the L phase region is much wider and extends to higher temperatures than in $2$D. However, details such as the shape of the phase boundaries, the size of the L phase region, or the width of the GL coexistence
region depend sensitively on the parameters of the model such as the colloid radius $R$ and the adsorption strength $\alpha$. In order to illustrate the sensitivity, in Fig.~\ref{fig:diff_par} we plot phase diagrams in the plane spanned by $(\eta, \Delta T/\Tcs)$ for different (fixed) reservoir concentrations $x_r$. This representation is useful in an experimental context. In this plot, the values of the parameters $R$ and $\alpha$ differ from those used previously. We see that for smaller particles and a composition $x_r$ close to the critical reservoir composition $x_{rc}=0.5$, panel (a), there is a wide region of L phase and GL coexistence persisting to $\Delta T=0$. Choosing reservoir compositions further from $x_{rc}$, the topology of the phase diagram changes quite dramatically: the GL critical point becomes metastable with respect to GS coexistence
and a GLS triple point emerges -see panel (b) and the L phase exists only very close to $\Delta T=0$. A similar trend was also observed in $2$D~\cite{edison2015}. Increasing the colloid size and adsorption strength makes the phase diagram even more complex- see panel (c). Now, at higher temperatures, an upper GLS triple point emerges in addition to the lower GLS triple point. Note again the small L region.

\subsection{Four-component mixture}
\label{subsec:2t}

In this sub-section, we present our results for the full four-component mixture, i.e. the binary solvent plus two types of colloid. Unless otherwise stated, we set the values of the parameters to be  $R = 5$ and $\alpha_1 = 0.29$, which we used to study the ternary mixture with only type 1 colloid. We consider several values of $\alpha_2$ to investigate how increasing the contrast in the strength with which both types of colloids adsorb solvent component B affects the phase behavior of the mixture.

\begin{figure}[h!]
     \includegraphics[width=\linewidth]{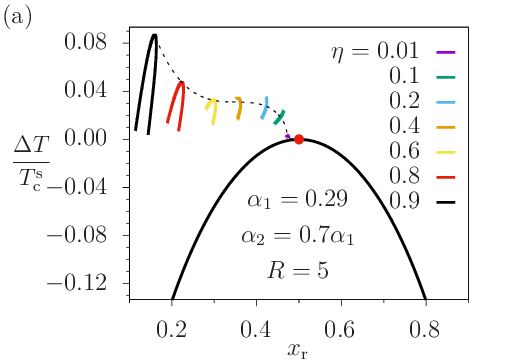}
     \includegraphics[width=\linewidth]{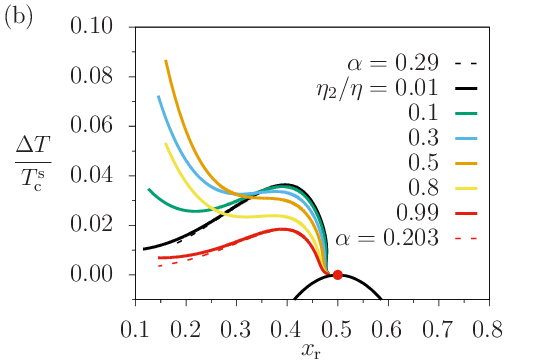}
    \caption{Demixing curve of the colloid free solvent (black line) and the critical lines for various fixed values of $\eta$ (a) and $\eta_2/\eta$ (b) and parameters $R=5$, $\alpha_1=0.29$, and $\alpha_2=0.7\alpha_1$. The dashed line in (a) corresponds to $\eta_2/\eta=0.5$. The dashed curves in (b) show  the limiting case of a ternary mixture with colloids of type 1 ( $\alpha=\alpha_1=0.29$)  and of type 2  ($\alpha=\alpha_2=0.203$). Recall that $x_\mathrm{r}$ is the reservoir composition.}
    \label{fig:Demix-TC-2-types-exp-cuts}
\end{figure}

In the previous sub-section we saw that the critical line obtained with the addition of colloids (see Fig.~\ref{fig:Demix-TC-1Type}) merges smoothly into the critical point of the pure solvent. Upon adding the second type of colloid, this line extends to a critical surface. Here we present cross-sections of this critical surface. We use experimentally useful cross-sections are those corresponding to either a fixed total colloid packing fraction $\eta$ or a fixed ratio of the amount of one type of colloid to the total number of colloids $\eta_2/\eta$.
The phase separation curve of the colloid free solvent and the critical lines for these two cases are shown in Fig.~\ref{fig:Demix-TC-2-types-exp-cuts} for $\alpha_1=0.29$ and $\alpha_2=0.7\alpha_1$. The critical lines in panel (a) correspond to different fixed values of $\eta$. Along these lines, the value of $\delta=\eta_2-\eta_1$ varies. The length of the lines is finite because $-\eta\le\delta\le \eta$, where for $\delta=-\eta(+\eta)$ there are only type 1 (type 2) colloids in the mixture. The larger the value of $\eta$, the longer the line of critical points in this representation. The lower panel of Fig.~\ref{fig:Demix-TC-2-types-exp-cuts} shows the critical point lines for various fixed values of $\eta_2/\eta$. The dashed lines correspond to the limiting cases of a mixture containing only type 1 colloids ($\alpha=\alpha_1=0.29$ at the top) and only type 2 colloids ($\alpha=0.7\alpha_1=\alpha_2=0.203$ at the bottom). For decreasing reservoir concentrations between the critical concentration and about  $x_\mathrm{r}\simeq 0.33$, the critical lines rise monotonically between the limiting cases increasing more rapidly with decreasing $\eta_2/\eta$. Note that these lines are solutions to the conditions given by (\ref{eq:crit_cond}) and for larger values of $\eta$ they become metastable with respect to S phase. 

\begin{figure}[htb!]
\includegraphics[width=\linewidth]{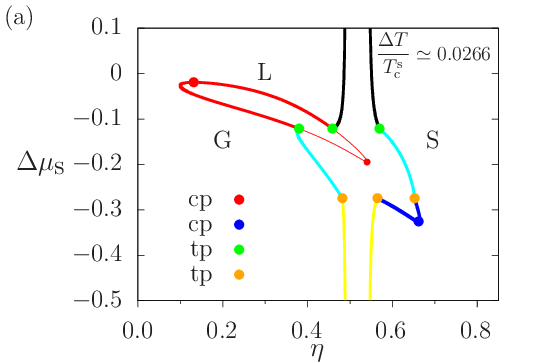}
\includegraphics[width=\linewidth]{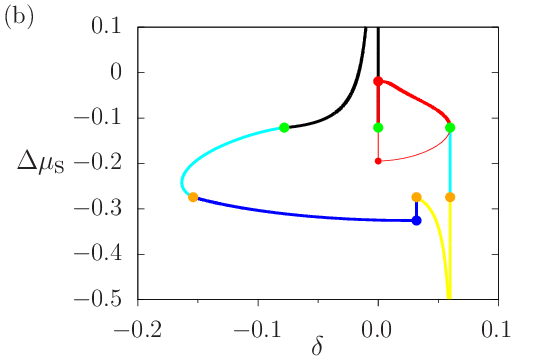}
    \caption{(a) Phase diagram in the ($\eta$, $\Delta \mu_\mathrm{s}$) plane at temperature $\Delta T/T_\mathrm{c}^\mathrm{s}=0.0266$ for $R=5$, $\alpha_1=0.29$ and $\alpha_2=0.9\alpha_1$. We set  $\delta =0$ in the F and L phases (black and red lines, respectively); see text. Along the G branch of GS
    coexistence (cyan and yellow  lines)  $\delta$ is fixed at its value in the G phase
    at the upper and lower triple points, (green and orange dots, respectively) ; see also  the vertical lines in panel (b).
    The GL transition (red curves) is metastable below the upper triple point as indicated by thin red lines ending at a (lower) critical point. (b) Same as in panel (a), but in the ($\delta$,$\Delta \mu_\mathrm{s}$) representation. The color code is the same. For any set of coexisting curves with the same color, the left curve in panel (a) corresponds to the right curve in panel (b) and vice versa- see text.}
    \label{fig:phasediagram-2types-same}
\end{figure}

We start to map out a phase diagram in the ($\eta$, $\Delta \mu_\mathrm{s})$ representation, see Fig.~\ref{fig:phasediagram-2types-same}, choosing a temperature close to that in Fig.~\ref{fig:low-T-4-fields} for the ternary mixture, in order to see what changes occur if we replace half of the colloids with colloids that attract the solvent slightly less ($\alpha_2=0.9\alpha_1$). To make this comparison, we fix the value of the parameter $\delta = \eta_2-\eta_1=0 $ in the F
phase above the GL critical point (red dot) and find coexistence with the S phase
(black line). Below the critical point there is also GL coexistence (red line); GLS coexistence occurs at the upper triple point (green dots). Note that the value of $\delta$ is equal to 0 along the F
branch of the black lines and along the  L branch
of the red lines, as described in the figure caption. This choice corresponds to the vertical black and red lines in panel (b) of Fig.~\ref{fig:phasediagram-2types-same}, where we represent the phase diagram in the  $(\delta, \Delta \mu_\mathrm{s})$ plane. In this panel we see that while in the L phase
we have the same fraction of each type of colloid, i.e. $\eta_1=\eta_2=1/2$ ($\delta=0$), in the coexisting S phase
the parameter $\delta<0$ indicating the presence in S of a larger fraction of colloidal type 1 particles which interact more strongly with the solvent. By contrast, in the coexisting G phase
we find $\delta>0$, indicating the presence of a larger fraction of colloidal type 2 that interact less strongly with the solvent. Below the upper triple point, where a stable L phase
ceases to exist, $\delta$ is fixed  at its value in the G phase at the triple point (cyan and yellow lines in Fig.~\ref{fig:phasediagram-2types-same}). This value is not zero: see vertical cyan and yellow lines in panel (b). Note that the orange dots indicate the lower triple point below which we observe GS coexistence (yellow lines) and SS coexistence (blue lines) ending at a (blue) critical point. It is important to note that we could have fixed the value of $\delta$ along other branches. Our present choice permits having continuous curves on the phase diagram. Comparing Fig.~\ref{fig:phasediagram-2types-same}(a) with Fig.~\ref{fig:low-T-4-fields} we see that for the selected parameter sets, the topology of the phase diagram is the same as for the ternary mixture. Indeed, changes, such as the position of critical and triple points or the size of coexistence regions, are very small. 

We emphasize that monitoring the parameter $\delta = \eta_2-\eta_1$ is crucial in ascertaining the complex nature of fractionation in these multicomponent mixtures. 

\begin{figure}
    \includegraphics[width=0.99\linewidth]{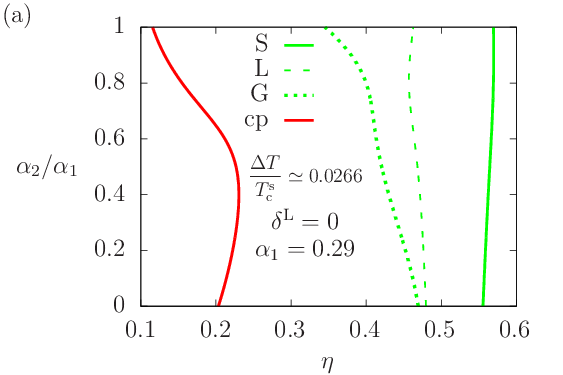}
    \includegraphics[width=0.99\linewidth]{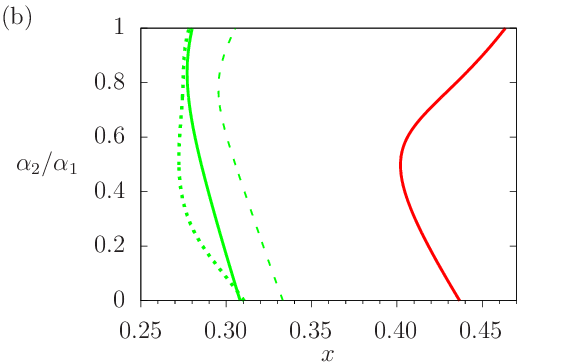}
   \includegraphics[width=0.99\linewidth]{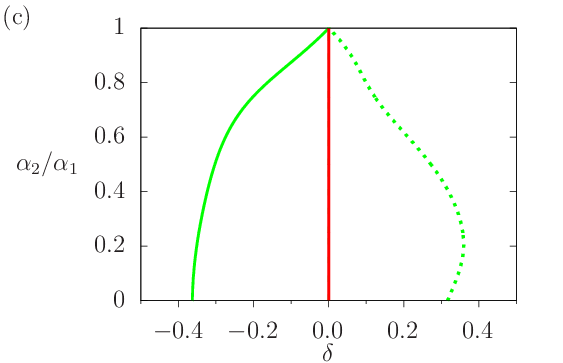}
    \caption{Effect of varying $\alpha_2$ on location of upper triple points (green lines) and on the GL critical point (red line). Panel (a) refers to the $(\eta,\alpha_2/\alpha_1)$ plane as in top panel of Fig.~\ref{fig:phasediagram-2types-same}. Panels (b) and (c) present the same results in the $(x,\alpha_2/\alpha_1)$ and $(\delta,\alpha_2/\alpha_1)$ planes. The color code is the same throughout.}
\label{fig:alpha2}
\end{figure}

Changes in phase boundaries become more pronounced as we increase the contrast between the adsorption preferences of the two colloidal types. We illustrate this in Fig.~\ref{fig:alpha2}. Panel (a) shows the shift in $\eta$ of i) the upper triple point (green curves) and ii) the stable upper critical point (red curve) from the results in Fig.~\ref{fig:phasediagram-2types-same} as we decrease $\alpha_2$ at a fixed temperature of $\Delta T/T_\mathrm{c}^\mathrm{s}=0.0266$ and fixed $\alpha_1=0.29$. The most pronounced variation with increasing contrast occur for the location of the GL critical point and for the G branch
of the triple point line (dotted green line). The corresponding S branch
of the triple line (solid green curve) is almost unchanged on varying  $\alpha_2$ ; this branch is dictated mainly by packing considerations, very similar to those for a monodisperse hard sphere system. Note that the value $\delta^\mathrm{L}=0$ is fixed in the L phase,
as in Fig.~\ref{fig:phasediagram-2types-same}.  Note also that the limit $\alpha_2/\alpha_1=1$ corresponds the case where the two colloids have identical adsorbing properties so that $\eta_2=\eta_1$ or $\delta=0$ for all phases, mimicking the situation in Fig.~\ref{fig:low-T-4-fields}.
This is clear in the bottom panel where the three (critical point, S and G branches of triple point) lines meet at $\alpha_2/\alpha_1=1$. Panel (b) displays the corresponding results with the fraction $x$ of the total number of sites occupied by type B colloids as the abscissa. 

\begin{figure}[htp!]
\centering
\includegraphics[width=8.3cm]{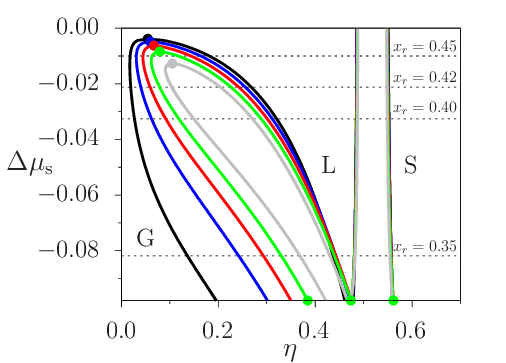}
\caption{  Phase diagram  at fixed temperature $\Delta T /T_\mathrm{c}^\mathrm{s}= 0.0103(3)$ in the ($\eta$,$\Delta\mu_s$) plane. Here  $R=5$, $\alpha_1=0.29$ and $\alpha_2=0.7\alpha_1$. Fixed ratios of $\eta_2/\eta$ in the L phase
are labeled black, blue, red, green, and gray, denoting $\eta_2/\eta=0,0.25,0.5,0.75,1$, respectively. GL transitions
terminate at the critical points (dots) shown in the upper left corner of the figure. The four dashed horizontal lines indicate different compositions $x_{\mathrm{r}}$ of solvent reservoir. The three green points on the $x$ axis label
G, L and S states coexisting at the (GLS) triple point for ratio $\eta_2/\eta$ = 0.75. The triple points for the other ratios lie below the range shown.}
  \label{fig:mf_1}
\end{figure}

A further way to represent the phase diagram for a mixture with two different colloidal types is to set the ratio $\eta_2/\eta=\mathrm{c}$, a constant, in one of the coexisting phases. This constraint sets $\delta=(2c-1)\eta$ in the corresponding phase. Following the same logic as in Fig.~\ref{fig:phasediagram-2types-same}, this constant value $c$ must be adjusted when the coexisting curves open out from triple points. An exemplar phase diagram is shown in Fig.~\ref{fig:mf_1} in the $\Delta\mu_s$ vs $\eta$ representation for fixed value of $\Delta T/T_\mathrm{c}^\mathrm{s}=0.010(3)$ and with $\alpha_1=0.29$ and $\alpha_2=0.7\alpha_1$. The overall topology of the phase diagram is the same as that in Fig.~\ref{fig:phasediagram-2types-same}. We focus here on  the range of $\Delta\mu_s$ that corresponds to the reservoir concentration  $x_\mathrm{r}$ of the solvent deviating slightly from its critical value $x_\mathrm{r,c}=0.5$, at which we might expect intuitively the critical Casimir effect to be strongest.  The different curves represent different ratios $\eta_2/\eta$, which are fixed  in the L phase.
It can be seen that the GL phase
boundary depends strongly on the relative amounts of type 1 and type 2 colloids.  As the ratio $\eta_2/\eta$ increases, recall type 2 has weaker attraction for the solvent, the critical value of $x_\mathrm{r}$ moves further away from $x_{r,c}=0.5$.

\begin{figure}[htp!]
\centering
\includegraphics[width=8.3cm]{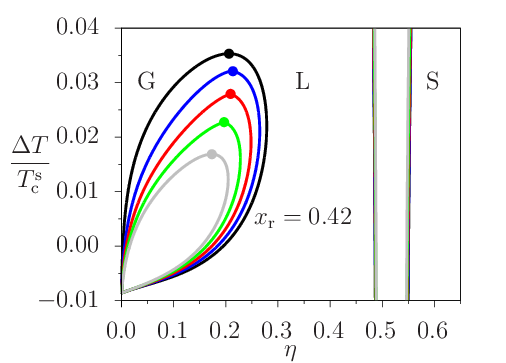}
\caption{Phase diagram at fixed reservoir concentration $x_{\mathrm{r}}$ in ($\eta, \Delta T/T^\mathrm{s}_\mathrm{c}$) plane. Here $R=5$, $\alpha_1=0.29$ and $\alpha_2=0.7\alpha_1$ as in Fig.~\ref{fig:mf_1} and we adopt the same color code to denote various fixed ratios of $\eta_2/\eta$ in the L phase.
Note i) how the GL critical point shifts to lower temperatures, i.e. towards $\Delta T =0$, as $\eta_2/\eta$ increases and ii) there is no triple point.}
\label{fig:9}
\end{figure}

In experiments, the phase behavior of the system is typically studied at a fixed solvent reservoir concentration $x_ r$. The variable parameters are temperature and the ratio of volume fractions  of the two types of colloids in the suspension. Figure~\ref{fig:9} shows the phase diagram for $x_ r = 0.42$ in the $(\eta, \Delta T/T^\mathrm{s}_\mathrm{c})$ plane.  For this particular choice of parameters, the phase diagram has a broad L phase
and for each ratio $\eta_2/\eta$ we observe GL coexistence persisting for temperatures down to  $\Delta T = 0$. Significantly we do not observe a triple point.

A more complex phase diagram topology arises upon increasing the radius $R$ of the colloids and the interaction parameter $\alpha_1$. We were motivated to look for this following the experimental study in Ref.~\cite{Kodger-et:2025}.

\begin{figure}[htb!]
        \includegraphics[width=\linewidth]{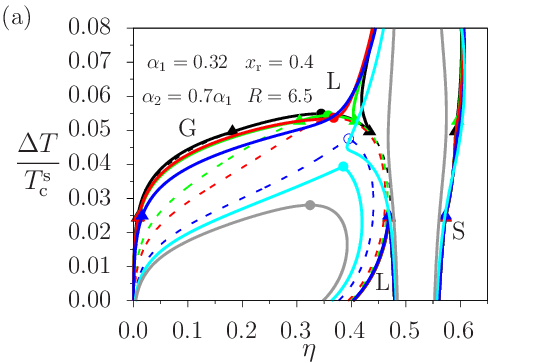}
               \includegraphics[width=\linewidth]{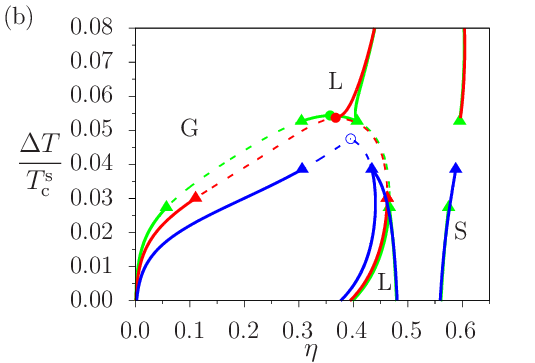}
        \caption{ {Phase diagram in ($\eta, \Delta T/T^\mathrm{s}_\mathrm{c}$) plane at $x_\mathrm{r}=0.4$ for various fixed values of $\eta_2/\eta$ (see below). Here $R=6.5$, $\alpha_1=0.32$ and $\alpha_2=0.7\alpha_1$.
        In both panels, the triangles denote GLS triple points and the full (open) circles are stable (metastable) critical points of GL transitions.
        The dashed curves denote metastable phase coexistence.  Black, green, red, blue, cyan, and gray lines correspond to  ratios $\eta_2/\eta=0,0.1, 0.187,0.5,0.73,1$, respectively, fixed in various phases. In panel (a) $\eta_2/\eta$ is fixed in the L phase for temperatures above the 
        upper and below the lower triple points (triangles) and in the G phase for temperatures between the triple points. In panel (b) $\eta_2/\eta$ is fixed in the L phase. Although the upper triple points are the same for both panels, the lower triple points differ due to the different scenarios taken in panels (a) and (b); see main text.}} 
     \label{fig:experiment-Big-R}
\end{figure}

In Fig.~\ref{fig:experiment-Big-R} we plot the phase diagram for $x_\mathrm{r}=0.4$, $R=6.5$, $\alpha_1=0.32$ and $\alpha_2=0.7\alpha_1$ for various values of $\eta_2/\eta$ fixed  in a way that  will be described below.  Panel (a) is a scenario similar to Fig.~\ref{fig:phasediagram-2types-same}, i.e., adapts the phase in which the ratio $\eta_2/\eta$ is fixed, whereas, panel (b) keeps the same phase (L) when fixing the ratio.

We first make some general remarks about this figure. In both panels the triangles are GLS triple points, full (open) circles denote stable (metastable) critical points of GL transitions and the dashed curves are metastable GL phase transitions. For high temperatures we have FS coexistence and in this temperature regime, we fix $\eta_2/\eta$ in the F phase in both panels. Upon reducing the temperature each system exhibits a GL critical point below which there is GL coexistence, denoted by a solid line if the GL transition is stable with respect to GS and by a dashed line if it is metastable. The colors correspond to choices of $\eta_2/\eta$; see caption.  If for a specific color code we do not encounter triple points we fix $\eta_2/\eta$ in the F phase to determine FS coexistence and in the L phase for GL coexistence.

The two panels differ when we encounter the GLS triple point. In panel (a) at temperatures between the upper triple point and the lower triple point the GL transition lies in the metastable regime and only the GS transition is stable. In order to have continuous curves between the triple points we fix $\eta_2/\eta$ in the G phase equal to its value at the upper triple point. This approach is similar to the one in Fig.~\ref{fig:phasediagram-2types-same}. At temperatures below the lower triple point we have again GL and LS transitions. In this regime we fix $\eta_2/\eta$ again in the L phase but this time equal to its value at the lower triple point.

Panel (b) is different from panel (a) in that we choose to fix $\eta_2/\eta$ in the L phase to be equal to the same value in the F phase at high temperature. Of course, in the temperature range between the triple points the GL transition is metastable. Importantly the upper triple points remain the same in both panels but the lower triple points in (b) are different from in panel (a). Recall that the (metastable) dashed curves are the same in both panels and that in panel (b) the lower triple points lie on these dashed curves which is not the case in panel (a). The convention in panel (b) is more appropriate for an experimental situation where a system is prepared with a fixed ratio of colloid fractions in a specific phase, for example a L phase.

The black and gray colors in panel (a) correspond to $\eta_2/\eta=0,1$ i.e., the limiting cases of pure colloid type 1 or 2, respectively. We observe that upon going from one to the other the topology of the phase diagram changes dramatically: the system of pure colloid type 1 (more strongly adsorbing of solvent species B) shows LS, GS and GL phase transitions, 
and upper and lower triple points, whereas the system of pure colloid type 2 shows LS and a disjoint GL phase transitions.

For small amounts of colloid type 2 (green curves) we have the same topology as in the black curve, showing a critical point (green circle) and two sets of triple points (green triangles). As described
above, in panel (a) we employ fixed values of $\eta_2/\eta$ at the triple points: above the upper triple points $\eta_2/\eta=0.1$ in the L phase, in the temperature range between between the triple points $\eta_2/\eta = 0.222$ in the G phase, corresponding to the value at the upper triple point, and below the lower triple point $\eta_2/\eta = 0.005$ in the L phase, corresponding to the value at the lower triple point. The dashed green curve corresponds to metastable LG phase transitions
with $\eta_2/\eta=0.1$ in the L phase.
The green curve in panel (b) shows the phase diagram if we insist on setting $\eta_2/\eta=0.1$ in the L phase. The upper part of the phase diagram, including the upper triple points, as well as the metastable LG transitions, marked by green dashed line, are the same as in panel (a). However, in panel (b) there is no stable L phase in the temperature range between the triple points so we do not show any phase boundary.

The topology of the phase diagram changes when a triple point disappears. The red curves illustrate this. Here the critical point of the GL transition
coexists with a S phase. GL coexistence at the same $\eta_2/\eta$ ratio in the L phase
is metastable (dashed red lines), and the system has lost its upper triple point. However, it still exhibits a lower triple point (red triangles), below which we must readjust the $\eta_2/\eta$ ratio in the L phase to be equal to its value at the triple point. In panel (a) $\eta_2/\eta \simeq 0.187$ in the F phase at higher temperatures and also for  GS until the temperature is lowered to the upper triple point (the red triangles). Insisting upon setting $\eta_2/\eta \simeq 0.187$ in the L phase
results in a different phase diagram, as shown in red in panel (b). Once again, the triple points differ from those in panel (a).

For $\eta_2/\eta=0.5$ (blue curves), the topology of the system remains the same as for the red curve. The only difference is that the GL critical point (open blue circle) lies inside the metastability region. Note again that the triple points with the same color differ between panels (a) and (b).

For $\eta_2/\eta \simeq0.697$ we find a GL critical point coexisting with the S phase, and the topology changes again: we lose the lower triple point, and the GL and LS transitions separate.
This new topology is represented by the cyan curve with $\eta_2/\eta=0.73$ in the L phase, both along the GL and LS phase boundaries.
 (Note that after the triple points disappear there is no need to maintain continuity at these points. Thus this topology is shown only in panel (a)). By adding more type-2 colloids to the mixture, the system approaches smoothly the other limiting case, i.e., the gray curve with $\eta_2/\eta=1$.

\begin{figure}[htb!]
        \includegraphics[width=\linewidth]{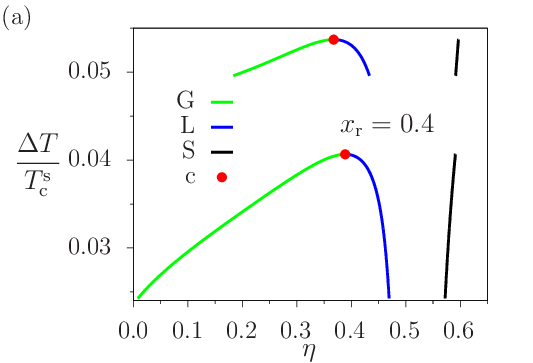}
        \includegraphics[width=\linewidth]{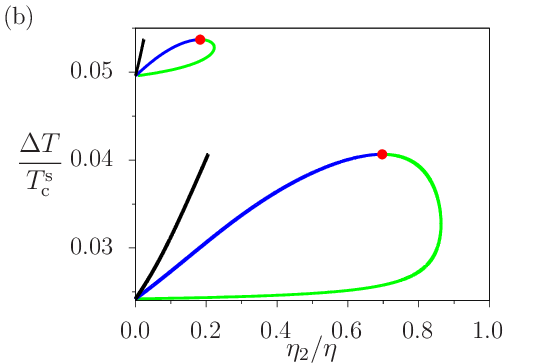}
       \caption{ (a) The loci of the triple points in the $(\eta,\Delta T/T_\mathrm{c}^\mathrm{s})$ plane. Here  $R=6.5$ and $\alpha_1=0.32$ and $\alpha_2=0.7\alpha_1$, as in Fig.~\ref{fig:experiment-Big-R}.
       The red points are GL critical points that coexist with the S phase. The green, blue, and black lines represent the
       G, L, and S phases, respectively,that coexist at a given temperature. (b) Results for the same parameters but now plotted in the $(\eta_2/\eta,\Delta T/T_\mathrm{c}^\mathrm{s})$ plane. The color code is the same as in panel (a). }
    \label{fig:upper-lower-riple-points}
\end{figure}

In order to expound how the topology of the phase diagrams depends on the appearance and disappearance of triple points, we plot the triple points in the  $(\eta,\Delta T/T^\mathrm{s}_\mathrm{c})$ and $(\eta_2/\eta,\Delta T/T^\mathrm{s}_\mathrm{c})$ planes in panels (a) and (b) of Fig.~\ref{fig:upper-lower-riple-points}. These results are for the same parameters as in Fig.~\ref{fig:experiment-Big-R}. The two branches of curves correspond to the upper and lower triple points. The two red points are the two GL critical points which coexist with a S phase.  The upper case is for $\eta_2/\eta \simeq 0.187$ in the L phase and corresponds to the red line in Fig.~\ref{fig:experiment-Big-R}, while the lower case corresponds to $\eta_2/\eta \simeq 0.697$ in the L phase. At these points, the G and L branches of the triple point disappear, which causes a change in the topology of the phase diagram.

\section{Summary and conclusions}
\label{sec:conclusion}

We have investigated the phase behavior of a dense mixture of two types of colloids
C$_\mathrm{1}$ and C$_\mathrm{2}$ suspended in a binary liquid solvent, employing a lattice model together with a mean-field treatment of the free energy of this four-component mixture. To some extent, our approach treats all species on equal footing unlike the picture that builds upon the assumption of effective pair potentials between colloids used in other studies~\cite{Vasilyev2009,Mohry-et:2012a,Mohry-et:2012b,Gnan-et:2012c,Mohry-et:2014,Dang-et:2013,Nguyen-et:2013,Marcel-et,maciolek2018collective}. Within our treatment both types of colloids are significantly larger than the other two components (binary solvent molecules) and both have an inherent preference for adsorbing species B of the solvent. For simplicity we choose the radius $R$ of both types to be identical. The only difference between the two types of colloids lies in the strength of their interaction, parameters $\alpha_i$, with solvent species B. In our calculations we chose to follow a typical experimental situation: the colloids are immersed in the single phase region of the binary solvent that is poor in the colloid-preferred species B, i.e. we chose reservoir compositions $x_\mathrm{r} < 0.5$ and temperatures $\Delta T >0$.
 We were motivated to extend earlier studies of the lattice model with a single colloidal type~\cite{Edison2015PRL,10.1063/1.4961437,edison2015,Tasios-et:2017} by recent experiments, aimed at addressing colloidal assembly, performed for mixtures of two types of colloids in a near critical binary solvent~\cite{Kodger-et:2025}. By changing the surface charge and therefore the solvent affinity of the colloidal particles it was possible to mix modified colloids with unmodified ones. Rich phase behavior was obtained with complex crystallization behavior reminiscent of that in atomic alloys. +
 Measurements were performed at constant solvent concentration and total colloid packing fraction, examining a large range of colloidal compositions at temperatures close to the solvent's critical temperature. Experimental observations along a typical cooling path, corresponding to a vertical line $\eta \simeq 0.4$ in Fig.~\ref{fig:experiment-Big-R}, are consistent with predictions from the present model and theory. Specifically, the predicted metastable GL transition and critical point are consistent with the observed GL density fluctuations and the predicted underlying GS coexistence is consistent with the stable GS coexistence observed at slightly lower temperatures.  Note that our predictions 
of additional critical and triple points have not been realized experimentally.
Looking forward, since the adsorption contrast between colloids type C$_1$ and C$_2$ can be tuned through surface charge the results from our model might guide how to adjust the physical parameters that determine the binary colloid phase diagram thereby enabling the exploration of pertinent 'binary alloy' properties.

 Returning to the gamut of results from the model, we find extremely rich phase behavior characterized by several parameters. This is not unexpected given what we know from studies with only one type of 
 colloid~\cite{Edison2015PRL,10.1063/1.4961437,edison2015,Tasios-et:2017}.
 What is particularly striking is that the topology of the phase diagram is very sensitive to the value of $\eta_2/\eta$. For small fractions of the less adsorbing colloids (type 2), we observe two regions of stable GL coexistence: a small upper GL coexistence at higher temperatures and a broad lower GL coexistence closer to the solvent's critical temperature. The upper GL coexistence disappears completely upon addition of more type 2 colloids to the suspension and is transformed into a broad region of GS coexistence. When the mixture consists mostly of weakly adsorbing colloids, a stable upper critical point reappears. On the other hand, the FS phase boundaries remain essentially invariant upon changing $\eta_2/\eta$ because they are mainly determined by packing considerations, resembling those for a one-component hard sphere system. 

Our model and the accompanying theory are minimal. The 'size ratios' we employ in calculations, i.e. the  parameter $R$, is not directly in keeping with experimental systems. Although our model incorporates (indirectly) a diverging correlation length of the binary solvent on approaching its critical point this at mean-field level. True critical Casimir effects are not incorporated. Whether these considerations are important in determining the overall phase behavior of dense colloidal suspensions remains to be ascertained.

Finally we note that whilst our model accounts for the contrast between the adsorption preferences, for species B of the binary solvent of the two colloidal types and thus the influence on phase behavior, it does not distinguish between structures of the solid phase, e.g.  which 'alloy crystal structure of colloidal particles ' might constitute a crystal with lowest free energy at a given temperature. Simulation studies would be valuable. Our present approach provides a guide or first step as to where in the plethora of physical parameters one might search for suitable choices for practicably useful colloidal mixtures.

\begin{acknowledgments}
The research of N.F.B and A.M. was funded by  the  National Science Center, Poland (Opus Grant  No.~2022/45/B/ST3/00936).
\end{acknowledgments}
\appendix
\section{Free energy of pure hard sphere colloidal system}
\label{sec: pure-colloids-calculations}

In this section we provide the explicit form for $\FC$, the free energy of the pure HS colloidal system.

We define $\psi$, the Helmholtz free energy per number of hard spheres as
$\psi:=\FC/\NC$. $\psi$ is a function of $T$ and $\rho:=\NC/V$, with volume $V=M$ in appropriate units.
In the fluid phase
\begin{equation}
\label{psi-decompose}
\psi^\text{F}(T,\rho)=\psi_\text{id}+\EXHS,
\end{equation}
where the index F denotes the fluid phase and
\begin{equation}
\label{id}
\psi_\text{id}(T,\rho)=\kB T\big(\ln(\Lambda^3\rho)-1\big),
\end{equation}
is the contribution from the ideal gas, where $\Lambda$ is  the de Broglie length and $\EXHS$ is the excess contribution from the hard sphere interactions which is described in the following. Note that we can replace $\rho$ by $\eta$
\begin{equation}
\label{etarho}
\eta=\frac{\frac{4}{3}\pi R^3\NC}{M}=\frac{4}{3}\rho\pi R^3=\rho\VC,
\end{equation}
In the fluid phase, we employ the accurate Carnahan Starling~\cite{carnahan1969} approximation:
\begin{equation}
\label{exhs}
\EXHS=\kB T\frac{\eta(4-3\eta)}{(1-\eta)^2}.
\end{equation}
Noting
\begin{equation}
\psi^\text{F}=\frac{\FC^\text{F}}{\NC}=\frac{\FC^\text{F}/M}{\rho}, 
\end{equation}
it follows
\begin{equation}
\label{F-fluid}
\begin{split}
\frac{\FC^\text{F}}{M}&=\rho\psi^\text{F}=\rho\psi_\text{id}+\rho\EXHS\\
&=\kB T\big(\frac{\eta}{\VC}\ln\Lambda^3\frac{\eta}{\VC}-\frac{\eta}{\VC}\big)
\\&
+\kB T\frac{\eta^2(4-3\eta)}{\VC(1-\eta)^2},
\end{split}
\end{equation}
with $\Lambda^3=\VC$.
Alternatively, using $\eta=\NC\VC/M$, we can write $\FC^\text{F}$ in terms of the volume as
\begin{equation}
\label{F-fluid-versus-M}
\begin{split}
\FC^\text{F}=&M\Big[\kB T\big(\frac{\NC}{M}\ln(\Lambda^3\frac{\NC}{M})-\frac{\NC}{M}\big)
\\&
+\kB T\frac{\VC\NC^2(4-3(\NC\VC/M))}{(M-\NC\VC)^2}\Big]\\
&=\kB T\NC\Big[\ln(\Lambda^3\frac{\NC}{M})-1
\\&
+\VC\NC\frac{4M-3\NC\VC}{(M-\NC\VC)^2}\Big].
\end{split}
\end{equation}
The reduced pressure in the fluid phase is given by the Carnahan--Starling result~\cite{carnahan1969}:
\begin{equation}
 \label{p-fluid}
 \frac{P^\text{F}}{\rho\kB T}=\frac{1+\eta+\eta^2-\eta^3}{(1-\eta)^3}.
\end{equation}
The above formulae for the free energy and pressure of hard spheres are valid up to the freezing point, which according to recent simulations~\cite{nayhouse2011} is $\eta_{\text{fr}}\simeq 0.4912$.
 At the fluid--solid phase transition two phases with two values of $\eta$ coexist; these two phases share the same value of pressure and chemical potential.

In order to find the free energy in the solid phase we use the results of Hall for the reduced pressure~\cite{hall1972}:
\begin{equation}
 \label{p-solid}
 \begin{split}
\Pi= \frac{P^\text{S}}{\rho\kB T}=&\frac{12-3\,\omega}{\omega}+2.557696+0.1253077\,\omega\\
 &+0.1762393\,\omega^2-1.053308\,\omega^3
 \\&
 +2.81862\,\omega^4-2.921934\,\omega^5
 \\&+1.118413\,\omega^6,
  \end{split}
\end{equation}
where the index S denotes the solid phase and
\begin{equation}
\label{omega-definition}
 \omega:=4(1-\frac{M_0}{M})=4(1-\frac{\eta}{\eta_0}),
\end{equation}
with $\eta_0=\eta_\text{close packing}=0.740480$.
Recall the pressures in the solid and fluid phases at coexistence must be identical, i.e. 
$P^\text{S}$ and $P^\text{F}$, at the freezing and melting point are the same.
The melting point from simulation~\cite{robles2014} is $\eta_{\text{m}}\simeq 0.543$.

In order to find the Helmholtz free energy of hard spheres in the solid phase we use:
\begin{equation}
\label{P-to-F-1}
\begin{split}
     P^\text{S}=\rho^2\frac{\partial \psi^\text{S}}{\partial\rho}
\end{split}
\end{equation} 
which implies
\begin{equation}
\label{P-to-F-2}
\begin{split}
    \psi^\text{S}=&\int \text{d}\rho\frac{P^\text{S}}{\rho^2}=\VC\int \text{d}\eta\frac{P^\text{S}}{\eta^2}
     \\&
     =\kB T\VC\int \text{d}\eta\frac{P^\text{S}}{\kB T}\frac{1}{\eta^2},
\end{split}
\end{equation}
where $P^\text{S}/(\kB T)=\Pi\,\eta/\VC$ is given by eq.~(\ref{p-solid}).
Equation~(\ref{P-to-F-2}) together with $\FC^\text{S}/M=\psi^\text{S}\rho=\psi^\text{S}\eta/\VC$ yields
the Helmholtz free energy as
\begin{equation}
\label{F-solid}
\begin{split}
\FC^\text{S}/M=&\frac{\kB T}{\VC}\eta\Big[O+2248.99\,\ln\eta-3\,\ln(0.740489-\eta)\\
&-20548.6\,\eta+39141.7\,\eta^2-52967.3\,\eta^3\\
&45297.6\,\eta^4 - 22003.8\,\eta^5 + 4631.26\,\eta^6\Big],
\end{split}
\end{equation}
where the constant $O$ comes from the integration in
eq.~(\ref{P-to-F-2}). In order to find the constant $O$ we first fix $\eta=\eta_{\text{fr}}\simeq 0.4912$ , the freezing value from  recent simulations~\cite{nayhouse2011}  and require that the
melting point $\eta_\text{m}$ is 
such that the pressure in the fluid and solid phases are the same ; note that the constant $O$ does not appear in the expression for the pressure. We find:
\begin{equation}
\begin{split}
 &P^\text{F}(\eta=\eta_{\text{fr}}\simeq 0.4912)=P^\text{S}(\eta)  
\end{split}
\end{equation}
which implies
\begin{equation}
\begin{split}
 &\eta=\eta_\text{m}\simeq 0.542455.   
\end{split}
\end{equation}
Now we ask for the value of the constant $O$ such that at this phase transition the chemical potentials are the same in the fluid and solid phase. The chemical potentials in the two phases are:
\begin{equation}
\label{muf}
\begin{split}
     \mu^\text{F}&=\VC\frac{\partial (\FC^\text{F}/M)}{\partial\eta}|_{T,M}
     \\&
     =\kB T\Big[ \frac{\eta (-8 + 9 \eta - 3 \eta^2)}{(-1 + \eta)^3} + 
    \ln\eta\Big],
\end{split}
\end{equation}
and
\begin{equation}
\label{mus}
\begin{split} 
 \mu^\text{S}&=\VC\frac{\partial (\FC^\text{S}/M)}{\partial\eta}|_{T,M}=
 \\&\frac{  \kB T }{-0.740489 + \eta} \Big[-1665.35 - 0.740489\,O\\
 &+ 32678\,\eta + 
 O\,\eta - 128049\,\eta^2 + 274312\,\eta^3\\
 &-379581\,\eta^4 + 324249\,\eta^5 - 156029\,\eta^6\\
 &+ 32418.8\,\eta^7 + (2.22147 - 3\,\eta)\,\ln(
     0.740489 - \eta)\\
     &+ (-1665.35 + 2248.99\,\eta)\,\ln\eta\Big],
\end{split}
\end{equation}
where in eq.~(\ref{muf}) we have put $\Lambda^3=\VC$.
Requiring the same chemical potentials at the freezing and melting point:
\begin{equation}
\label{constant-O-pre}
\begin{split}
& \mu^\text{F}(\eta_{\text{fr}}\simeq 0.4912)=\mu^\text{S}(\eta_{\text{m}}\simeq0.542455),
\end{split}
\end{equation}
implies that the constant
\begin{equation}
\label{constant-O}
\begin{split}
& O=6451.95.    
\end{split}
\end{equation}
Thus, the Helmholtz free energy is
given in three parts:
$\FC^\text{F}/M$ for $\eta<\eta_{\text{fr}}$,
$\FC^\text{S}/M$ for $\eta>\eta_{\text{m}}$, and the  straight line 
connecting these two curves between $\eta=\eta_{\text{fr}}$ and
$\eta=\eta_{\text{m}}$.

It is important to mention that when we add the solvent to the system for the full model (see eq.~\ref{F-Final}), the free energy of colloids $\FC^\text{F}$ and $\FC^\text{S}$ stay the same as calculated for the pure colloid case. This
means that the constant $O$ in $\FC^\text{S}$ given by eq.~(\ref{constant-O}) stays the same and is written explicitly in eq.~(\ref{F-solid-In-Text})

\bibliographystyle{unsrt}
\bibliography{references}

\begin{thebibliography}{10}

\bibitem{Kodger-et:2025}
T.~E. Kodger, N.~Farahmand~Bafi, M.~Labb\'e-Laurent, E.~Steijlen, A.~Macio\l~ek, , and P.~Schall.
\newblock Composite colloidal assembly by critical casimir forces.
\newblock {\em https://doi.org/10.48550/arXiv.2602.12431}, 2026.

\bibitem{doi:10.1021/acs.chemrev.6b00196}
Michael~A. Boles, Michael Engel, and Dmitri~V. Talapin.
\newblock Self-assembly of colloidal nanocrystals: From intricate structures to functional materials.
\newblock {\em Chemical Reviews}, 116(18):11220--11289, 2016.
\newblock PMID: 27552640.

\bibitem{doi:10.1021/ja3097527}
Yijin Kang, Xingchen Ye, Jun Chen, Yun Cai, Rosa~E. Diaz, Radoslav~R. Adzic, Eric~A. Stach, and Christopher~B. Murray.
\newblock Design of pt–pd binary superlattices exploiting shape effects and synergistic effects for oxygen reduction reactions.
\newblock {\em Journal of the American Chemical Society}, 135(1):42--45, 2013.
\newblock PMID: 23214936.

\bibitem{doi:10.1073/pnas.1422649112}
Daniel~J. Park, Chuan Zhang, Jessie~C. Ku, Yu~Zhou, George~C. Schatz, and Chad~A. Mirkin.
\newblock Plasmonic photonic crystals realized through dna-programmable assembly.
\newblock {\em Proceedings of the National Academy of Sciences}, 112(4):977--981, 2015.

\bibitem{Pusey1989}
P.~N. Pusey and W.~van Megen.
\newblock Phase behaviour of concentrated suspensions of nearly hard colloidal spheres.
\newblock {\em Nature}, 320:340--342, 1989.

\bibitem{Bartlett1992}
P.~Bartlett, R.~H. Ottewill, and P.~N. Pusey.
\newblock Freezing of binary mixtures of colloidal hard spheres.
\newblock {\em Phys. Rev. Lett.}, 68:3801--3804, 1992.

\bibitem{Leunissen2005}
M.~E. Leunissen, C.~G. Christova, A.-P. Hynninen, C.~P. Royall, A.~I. Campbell, A.~Imhof, M.~Dijkstra, R.~van Roij, and A.~van Blaaderen.
\newblock Ionic colloidal crystals of oppositely charged particles.
\newblock {\em Nature}, 437:235--240, 2005.

\bibitem{Solomon2010}
M.~J. Solomon and P.~T. Spicer.
\newblock Microstructural regimes of colloidal rod suspensions, gels, and glasses.
\newblock {\em Soft Matter}, 6:1391--1400, 2010.

\bibitem{Lu2008}
P.~J. Lu, E.~Zaccarelli, F.~Ciulla, A.~B. Schofield, F.~Sciortino, and D.~A. Weitz.
\newblock Gelation of particles with short-range attraction.
\newblock {\em Nature}, 453:499--503, 2008.

\bibitem{Floter1995}
G.~Fl{\"o}ter and S.~Dietrich.
\newblock Critical casimir forces in colloidal suspensions.
\newblock {\em Z. Phys. B}, 97:213--225, 1995.

\bibitem{Fisher1978}
M.~E. Fisher and P.~G. de~Gennes.
\newblock Phenomena at the walls in a critical binary mixture.
\newblock {\em C. R. Acad. Sci. Paris B}, 287:207--209, 1978.

\bibitem{schmidt2023tunable}
Falko Schmidt, Agnese Callegari, Abdallah Daddi-Moussa-Ider, Battulga Munkhbat, Ruggero Verre, Timur Shegai, Mikael K{\"a}ll, Hartmut L{\"o}wen, Andrea Gambassi, and Giovanni Volpe.
\newblock Tunable critical casimir forces counteract casimir--lifshitz attraction.
\newblock {\em Nature Physics}, 19(2):271--278, 2023.

\bibitem{D0NR09076J}
Oleg~A. Vasilyev, Emanuele Marino, Bas~B. Kluft, Peter Schall, and Svyatoslav Kondrat.
\newblock Debye vs. casimir: controlling the structure of charged nanoparticles deposited on a substrate.
\newblock {\em Nanoscale}, 13:6475--6488, 2021.

\bibitem{wang2024nanoalignment}
Gan Wang, Piotr Nowakowski, Nima Farahmand~Bafi, Benjamin Midtvedt, Falko Schmidt, Agnese Callegari, Ruggero Verre, Mikael K{\"a}ll, Siegfried Dietrich, Svyatoslav Kondrat, et~al.
\newblock Nanoalignment by critical casimir torques.
\newblock {\em Nature communications}, 15(1):5086, 2024.

\bibitem{10.1063/5.0235449}
Piotr Nowakowski, Nima Farahmad~Bafi, Giovanni Volpe, Svyatoslav Kondrat, and S.~Dietrich.
\newblock Critical casimir levitation of colloids above a bull’s-eye pattern.
\newblock {\em The Journal of Chemical Physics}, 161(21):214114, 12 2024.

\bibitem{Hertlein-et:2008}
C.~Hertlein, L.~Helden, A.~Gambassi, S.~Dietrich, and C.~Bechinger.
\newblock Direct measurement of critical casimir forces.
\newblock {\em Nature}, 451:172, January 2008.

\bibitem{Bonn-et:2010a}
Daniel Bonn, Gerard Wegdam, and Peter Schall.
\newblock Bonn, wegdam, and schall reply:.
\newblock {\em Phys. Rev. Lett.}, 105:059602, Jul 2010.

\bibitem{Gambassi-et:2009}
A.~Gambassi, A.~Macio\l{}ek, C.~Hertlein, U.~Nellen, L.~Helden, C.~Bechinger, and S.~Dietrich.
\newblock Critical casimir effect in classical binary liquid mixtures.
\newblock {\em Phys. Rev. E}, 80:061143, 2009.

\bibitem{Marcel-et}
S.~G. Stuij, M.~Labbe-Laurent, T.~E. Kodger, A.~Maciolek, and P.~Schall.
\newblock Critical casimir interactions between colloids around the critical point of binary solvents.
\newblock {\em Soft Matter}, 13:5233--5249, 2017.

\bibitem{maciolek2018collective}
Anna Macio{\l}ek and Siegfried Dietrich.
\newblock Collective behavior of colloids due to critical casimir interactions.
\newblock {\em Reviews of Modern Physics}, 90(4):045001, 2018.

\bibitem{Soyka-et:2008}
Florian Soyka, Olga Zvyagolskaya, Christopher Hertlein, Laurent Helden, and Clemens Bechinger.
\newblock Critical casimir forces in colloidal suspensions on chemically patterned surfaces.
\newblock {\em Phys. Rev. Lett.}, 101:208301, Nov 2008.

\bibitem{Troendle-et:2011}
M.~Tr{\"o}ndle, O.~Zvyagolskaya, A.~Gambassi, A.~Vogt, L.~Harnau, C.~Bechinger, and S.~Dietrich.
\newblock Trapping colloids near chemical stripes via critical casimir forces.
\newblock {\em Mol. Phys.}, 109:1169, 2011.

\bibitem{Beysens-et:1985}
D.~Beysens and D.~Est\`eve.
\newblock Adsorption phenomena at the surface of silica spheres in a binary liquid mixture.
\newblock {\em Phys. Rev. Lett.}, 54:2123, 1985.

\bibitem{MARINO2016154}
Emanuele Marino, Thomas~E. Kodger, Jan~Bart ten Hove, Aldrik~H. Velders, and Peter Schall.
\newblock Assembling quantum dots via critical casimir forces.
\newblock {\em Solar Energy Materials and Solar Cells}, 158:154--159, 2016.

\bibitem{Iwashita-et:2014}
Yasutaka Iwashita and Yasuyuki Kimura.
\newblock Orientational order of one-patch colloidal particles in two dimensions.
\newblock {\em Soft Matter}, 10:7170--7181, 2014.

\bibitem{Guo-et:2008}
Hua Guo, Theyencheri Narayanan, Michael Sztuchi, Peter Schall, and Gerard~H. Wegdam.
\newblock Reversible phase transition of colloids in a binary liquid solvent.
\newblock {\em Phys. Rev. Lett.}, 100:188303, May 2008.

\bibitem{Bonn-et:2009}
Daniel Bonn, Jakub Otwinowski, Stefano Sacanna, Hua Guo, Gerard Wegdam, and Peter Schall.
\newblock Direct observation of colloidal aggregation by critical casimir forces.
\newblock {\em Phys. Rev. Lett.}, 103:156101, Oct 2009.

\bibitem{Nguyen-et:2013}
V.~D. Nguyen, Suzanne Faber, Zhibing Hu, Gerard~H. Wegdam, and Peter Schall.
\newblock Controlling colloidal phase transitions with critical casimir forces.
\newblock {\em Nature Comm.}, 4:1584, 2013.

\bibitem{Zvyagolskaya-et:2011}
O.~Zvyagolskaya, A.~J. Archer, and C.~Bechinger.
\newblock Criticality and phase separation in a two-dimensional binary colloidal fluid induced by the solvent critical behavior.
\newblock {\em EPL (Europhysics Letters)}, 96(2):28005, 2011.

\bibitem{Vasilyev2009}
O.~Vasilyev, A.~Gambassi, A.~Macio{\l}ek, and S.~Dietrich.
\newblock Universal scaling functions of critical casimir forces obtained by monte carlo simulations.
\newblock {\em Phys. Rev. E}, 79:041142, 2009.

\bibitem{Mohry-et:2012a}
T.~F. Mohry, A.~Macio\l~ek, and S.~Dietrich.
\newblock Phase behavior of colloidal suspensions with critical solvents in terms of effective interactions.
\newblock {\em J. Chem. Phys.}, 136(22):224902, 2012.

\bibitem{Mohry-et:2012b}
T.~F. Mohry, A.~Macio\l~ek, and S.~Dietrich.
\newblock Structure and aggregation of colloids immersed in critical solvents.
\newblock {\em J. Chem. Phys.}, 136(22):224903, 2012.

\bibitem{Gnan-et:2012c}
Nicoletta Gnan, Emanuela Zaccarelli, and Francesco Sciortino.
\newblock Tuning effective interactions close to the critical point in colloidal suspensions.
\newblock {\em J. Chem. Phys.}, 137(8):084903, 2012.

\bibitem{Mohry-et:2014}
T.~F. Mohry, S.~Kondrat, A.~Maciolek, and S.~Dietrich.
\newblock Critical casimir interactions around the consolute point of a binary solvent.
\newblock {\em Soft Matter}, 10:5510--5522, 2014.

\bibitem{Dang-et:2013}
Minh~Triet Dang, Ana~Vila Verde, Van~Duc Nguyen, Peter~G. Bolhuis, and Peter Schall.
\newblock Temperature-sensitive colloidal phase behavior induced by critical casimir forces.
\newblock {\em J. Chem. Phys.}, 139(9):094903, 2013.

\bibitem{Sluckin:1990}
T.~J. Sluckin.
\newblock Wetting phenomena and colloidal aggregation in binary mixtures.
\newblock {\em Phys. Rev. A}, 41:960, 1990.

\bibitem{Loewen:1995}
H.~L{\"o}wen.
\newblock Solvent-induced phase separation in colloidal fluids.
\newblock {\em Phys. Rev. Lett.}, 74:1028, 1995.

\bibitem{GIL1998245}
Tamir Gil, John~Hjort Ipsen, Ole~G Mouritsen, Mads~C Sabra, Maria~M Sperotto, and Martin~J Zuckermann.
\newblock Theoretical analysis of protein organization in lipid membranes.
\newblock {\em Biochimica et Biophysica Acta (BBA) - Reviews on Biomembranes}, 1376(3):245--266, 1998.

\bibitem{Edison2015PRL}
John~R. Edison, Nikos Tasios, Simone Belli, Robert Evans, Ren\'e van Roij, and Marjolein Dijkstra.
\newblock Critical casimir forces and colloidal phase transitions in a near-critical solvent: A simple model reveals a rich phase diagram.
\newblock {\em Phys. Rev. Lett.}, 114:038301, Jan 2015.

\bibitem{10.1063/1.4961437}
Nikos Tasios, John~R. Edison, René van Roij, Robert Evans, and Marjolein Dijkstra.
\newblock Critical casimir interactions and colloidal self-assembly in near-critical solvents.
\newblock {\em The Journal of Chemical Physics}, 145(8):084902, 08 2016.

\bibitem{edison2015}
John~R. Edison, Simone Belli, Robert Evans, René van Roij, and Marjolein Dijkstra.
\newblock Phase behaviour of colloids suspended in a near-critical solvent: a mean-field approach.
\newblock {\em Molecular Physics}, 113(17--18):2546--2555, 2015.

\bibitem{Tasios-et:2017}
Nikos Tasios and Marjolein Dijkstra.
\newblock From 2d to 3d: Critical casimir interactions and phase behavior of colloidal hard spheres in a near-critical solvent.
\newblock {\em J. Chem. Phys.}, 146(13):134903, 2017.

\bibitem{carnahan1969}
Norman~F. Carnahan and Kenneth~E. Starling.
\newblock Equation of state for nonattracting rigid spheres.
\newblock {\em The Journal of Chemical Physics}, 51(2):635--636, 1969.

\bibitem{Speedy1998}
R.~J. Speedy.
\newblock The hard sphere equation of state revisited.
\newblock {\em J. Phys.: Condens. Matter}, 10:4387--4394, 1998.

\bibitem{robles2014}
Miguel Robles, Mariano~López de~Haro, and Andrés Santos.
\newblock Note: Equation of state and the freezing point in the hard-sphere model.
\newblock {\em The Journal of Chemical Physics}, 140(13):136101, 2014.

\bibitem{nayhouse2011}
Michael Nayhouse, A.~M. Amlani, and George Orkoulas.
\newblock A monte carlo study of the freezing transition of hard spheres.
\newblock {\em Journal of Physics: Condensed Matter}, 23(32):325106, 2011.

\bibitem{hall1972}
Kenneth~R. Hall.
\newblock Another hard‐sphere equation of state.
\newblock {\em The Journal of Chemical Physics}, 57(6):2252--2254, 1972.

\bibitem{tester1997thermodynamics}
Jefferson~W. Tester and Michael Modell.
\newblock {\em Thermodynamics and Its Applications}.
\newblock Prentice Hall, Upper Saddle River, NJ, 3rd edition, 1997.

\end{thebibliography}

\end{document}